# Progenetix: 12 years of oncogenomic data curation

Haoyang Cai, Nitin Kumar, Ni Ai, Saumya Gupta, Prisni Rath and Michael Baudis*

Institute of Molecular Life Sciences and Swiss Institute of Bioinformatics, University of Zurich, Winterthurerstrasse 190, 8057 Zurich, Switzerland

## ABSTRACT

DNA copy number aberrations (CNAs) can be found in the majority of cancer genomes, and are crucial for understanding the potential mechanisms underlying tumor initiation and progression. Since the first release in 2001, the Progenetix project (http://www.progenetix.org) has provided a reference resource dedicated to provide the most comprehensive collection of genome-wide CNA profiles. Reflecting the application of comparative genomic hybridization (CGH) techniques to tens of thousands of cancer genomes, over the past 12 years our data curation efforts have resulted in a more than 60-fold increase in the number of cancer samples presented through Progenetix. In addition, new data exploration tools and visualization options have been added. In particular, the gene specific CNA frequency analysis should facilitate the assignment of cancer genes to related cancer types. Additionally, the new user file processing interface allows users to take advantage of the online tools, including various data representation options for proprietary data pre-publication. In this update article, we report recent improvements of the database in terms of content, user interface and online tools.

* To whom correspondence should be addressed; Email: michael.baudis@imls.uzh.ch
Correspondence may also be addressed to Haoyang Cai, Email: haoyang.cai@gmail.com

# INTRODUCTION

DNA copy number aberrations are a form of genomic mutations found in the majority of individual cancer genomes (1-3). Most cancer types, especially solid tumors, exhibit distinct patterns of CNAs that may reveal both shared and distinct evolutionary processes in the development of different tumor entities (4-6). Understanding the role CNAs play in cancer initiation and progression should help to elucidate these mechanisms of oncogenesis (2,7). A subset of genomic rearrangements involves distinct oncogenes and tumor suppressors, either through the alteration of gene expression profiles or through the formation of oncogenic fusion genes, and directly promote cancer growth and metastasis (8-10). In clinical research, CNAs have been successfully employed to distinguish cancer subtypes and also been recognized as prognostic markers, with potential applications in therapeutic stratification (11,12).

Comparative genomic hybridization (CGH) is a class of *in situ* hybridization techniques that has extensively been used to screen genome-wide copy number aberrations in cancer samples (13,14). According to the different substrates, CGH platforms can be divided into chromosomal CGH (cCGH) and variants of array CGH (aCGH) (15-17). We apply the term "array CGH" broadly to cover all types of arrays resulting in whole genome copy number status data, including genomic single color arrays (e.g. SNP arrays), for which external reference data is used. For cCGH, normal metaphase chromosomes from cultured cells are used as hybridization target (13). In aCGH platforms, an array of defined DNA fragments is either spotted on a substrate (i.e. glass slide) through a variety of "printing" techniques, or is generated through *in situ* synthesis of DNA oligonucleotides (15,16). For all types of hybridization targets, genomic DNA extracted from a tumor sample is fluorescencelabeled and hybridized to the denatured target DNA. For dual-color experiments (e.g. cCGH, large insert clone arrays), a co-hybridization with normal genomic reference DNA labeled with a different fluorochrome is performed; variations in the tumor/normal fluorescence intensity ratios allow the detection of abnormal genomic content in the tumor at the corresponding genome loci (18-20). Single color array experiments require a computational evaluation of the signal distribution in relation to external reference datasets (21,22). Generally, the resolution of cCGH is limited to chromosome-bands level, and only genomic gains and losses greater than approximately 5-10 Mb can be reliably detected by cCGH (13). For aCGH, the resolution is determined by the number and size of probes on the array (23). Recently, several ultra- high resolution aCGH platforms have been manufactured with millions of probes on a single glass chip with the ability to detect minute genomic aberrations, as small as a few kb (24,25). These platforms include array types originally designed for other purposes, such as single nucleotide polymorphism (SNP) arrays and DNA methylation arrays (17,26).

The development of CGH and related techniques has greatly stimulated interest in cytogenetic analysis of different cancers (4,27-29). In the last decade, numerous oncogenomic data sets have been accumulated, making large-scale analysis across multiple cancer entities feasible. For instance, while our group previously had provided a descriptive analysis of more than 5000 CGH profiles from epithelial neoplasias (30), Beroukhim reported a study of copy number profiles of more than 3000 cancer samples, mainly from 26 entities (31). In 2013, Kim presented an analysis based on about 8000 genomic arrays (32). These analyses exemplified the value of large-scale CNA data analysis in cancer research. Furthermore, the comprehensive collection of genomic copy number profile scan be used to explore relatively low frequency gene-specific CNAs as well as complex events in cancer genomes, such as chromothripsis-like patterns (33,34).

Given the large amount of CGH data scattered in publications and various data repositories, it is highly desirable to have a single, comprehensive and well-annotated cancer CNA data resource. The Progenetix database aims to provide this kind of service to the research community (35). Chromosome and array CGH data in the full-text or supplementary files of published papers are extracted, processed and stored in the database in a standard format (36). Increasingly, data processed from raw probe files as part of the arrayMap project is added after supervised analysis and data review (37). In contrast to arrayMap with its representation of probe intensity data, Progenetix captures the robust, qualitative aspects of CNA (mapping and directionality), without attempts towards fine-grained interpretation of CNA magnitude or interpretation of absolute copy numbers.

Other resources for annotated CGH data include NCBI SKY/M-FISH and CGH (38), The Cancer Genome Atlas (TCGA) (29), CanGEM (39), CaSNP (40) and arrayMap (37). Each of these databases focuses on particular data sources or techniques, and provides unique features. So far, the Progenetix database represents the quantitatively largest resource for annotated CNA data from whole genome profiling experiments in cancer (35). As a result of over 12 years of data curation, the database currently contains over 30000 individual CNA profiles from several hundreds of cancer types. Moreover, it provides a wealth of associated clinical information curated from publication texts or supplementary data files.

Here, we describe recent feature updates and relevant improvements of the Progenetix resource, and demonstrate the novel data visualization interface and online analysis tools that have been added since the database was released 12 years ago.

## 12-YEAR DATA GROWTH

At the time of the original publication of the Progenetix resource in 2001, our website represented a first effort to provide a single resource for accessing whole genome copy number profiling data from CGH experiments (35). The database contained a total of 490 cases collected from 19 publications. The basic inclusion criteria were a) whole genome CNA data from b) cancer or pre-malignant samples and c) presented in peer-reviewed publications. In contrast to other resources, e.g. the SKY/M-FISH and CGH database then under preparation at the NCBI (38,41), Progenetix was intentionally designed as a curated database without user driven upload and data manipulation options. Although many quantitative and qualitative improvements having been implemented over the years, these basic design decisions have remained in place.

The latest release of Progenetix (July 2013) now presents 30687 samples from 1006 publications, representing a more than 60-fold data expansion compared to the resources initial state. Each sample presents the whole-genome copy number profile of an individual specimen (DNA from a cancer or leukemia sample or cell line). Included in the data set are 10261 CNA profiles generated by genomic array experiments, while the remaining are based on chromosomal CGH. Currently, samples in Progenetix have been classified into 363 cancer types according to the International Classification of Diseases in Oncology, 3rd Edition (ICD-O 3) (42). Table 1 lists the summary of database content classified by disease locus.

The dramatic increase in data content is primarily the result of the continuous expansion of published studies containing CGH based data. To ensure a complete identification of articles, we rely on a complex combination of keywords to search PubMed and evaluate returned as well as referenced articles with respect to data fromCGH analyses on cancer samples. Before the subsequent data extraction step, these studies must fulfill two basic criteria: i) The results are obtained from a complete whole-genome screening (with or without sex chromosomes) and ii) Experiments were on a case-by-case basis (i.e., no pooled samples). So far, we have been able to identify 2390 such publications, reporting 35703 chromosomal CGH and 68546 array based CNA profiling experiments. As the survey data indicates, only a minority of all identified articles contain accessible case-level data. Frequently, the authors provide only summary results or describe the copy number alterations in selected genes or regions of interest, instead of providing the sample specific whole genome CNA data generated through their experiments.

If sample specific CNA data is available, a variety of formats can be encountered, such as plain text description in ISCN (43) related formats or based on "Golden Path" coordinates. Data in supplementary materials of array based studies is frequently given as probe specific normalized log2 ratios. For the sake of convenience of storage and analysis of CGH data from different formats, a data processing and collection pipeline was established with the final format being the "Golden Path" mapped copy number status information of imbalanced genomic regions. ISCN style data is converted using a dedicated, regular expression based engine, while for array probe based data sets (e.g. raw CEL files or log2 value tables) standard segmentation and thresholding methods are being employed (for details pls. refer to (37)).

Besides the absolute content of the database with respect to the number of individual records, the information depth of associated data has been extended greatly. While originally clinical data was limited to the consistently available diagnostic classification and sample locus information, we recently put emphasis on extracting other types of clinical data with possible relevance for the association with genomic features. Publications were inspected by the database curators to extract case-level clinical features, including patient age, gender, follow-up, survival status, tumor stage, grade, and sample source (e.g. primary vs. metastasis, recurrence, cell line). Although these criteria are not consistently available, the vast amount of samples will allow for integration of these associated features in analyses of considerable data sets; e.g., the current edition contains 3853 samples with complete follow-up/survival data (Supplementary Figure S1).

## USER INTERFACE

Representation of copy number aberration data in Progenetix is based on the principles of a) aggregation of CNA data for different classification values, and b) active aggregation of CNA data for dynamically generated data subsets. Pre-defined data categories with automatic one-click data aggregation are publications (defined by PMIDs), ICD-O 3 morphologies (ICD-O 3 codes), disease loci (ICD topography codes) and SEER (Surveillance, Epidemiology and End Results) categories (44). Additionally, for the majority of cases we have assigned a value for a category called "Clinical Groups" (see below).

Of the different categories, publications have a special place since we also present publication entries for articles with discussion of CGH data sets but without corresponding samples in the database (e.g. no sample specific data is listed or deposited). The "Search Publications" page contains the following input fields: (i) Author Name; (ii) Title Keyword: search for keyword occurrence in publication titles; (iii) Text Search: search for keyword occurrence in title, author, journal and abstract; (iv) PubMed IDs. In addition, filtering is available to (i) select only publications with data in Progenetix and (ii) to limit publications to those containing cases generated by aCGH, cCGH or both platforms.

### Active data aggregation

While earlier editions of the Progenetix resource relied on fixed categories with pre-computed CNA frequency profiles, the current version is based on sample specific data with dynamic search and aggregation options. Samples can either be retrieved using *ab novo* queries, or can be selected based on the categories above. Each option leads to a second selection step, in which all values existing in the currently active data subset are presented for an extended list of categories, allowing for further selection-based restriction of the data before processing and visualization.

For the *ab novo* data retrieval, the "Search Samples" form has options to query for: (i)

Text: free text search over most fields; (ii) ICD-O-3 Code: full or partial (start-anchored); (iii) ICD Topography Code: full or partial; (iv) PMID; (v) Sample IDs; (vi) Array Series IDs, e.g. GEO "GSE"; (vii) Sample Source: metastasis, cell line and primary tumor. Other selectors include "Technique" (aCGH vs. cCGH) and the option to only display data with completed clinical follow-up. All these search fields can be combined using boolean AND (intersection) or boolean OR (union) mode.

For the new version of Progenetix, one noteworthy feature is given through "Find CNAs by Gene or Region", which is particularly useful for gene specific queries. Gene names, chromosome bands or regions of interest can be specified, and tumor samples with CNAs in the queried genomic regions will be returned. Such analysis may be able to pinpoint cancer genes that are disturbed and may have a causative relationship to the corresponding cancer. The input box suggests plausible gene names and supports auto-completion. Moreover, the type of copy number alterations, gain, loss or both, can be specified.

As an example usage scenario, here we use the interface to explore data related to carcinomas of the esophagus. We start by using the keyword "esophagus" in the "Text Search". 475 samples are returned and presented on the "Sample Selection" page (Figure 1A). In this page, the search results can further be filtered to present only samples fulfilling selected criteria. Clicking on the "Sample Details" button will present the list of all samples with detailed information, such as clinical data and links to PubMed and/or GEO datasets, where available. In the following section of the page, several selector fields provide the corresponding values encountered in these 475 samples. Items in these blocks can be selected for the next analysis step. In this example, categories include: (i) Article: the samples were derived from 22 publications; (ii) Morphology: 8 ICD-O-3 types are displayed; (iii) Locus: 2 tumor sites are presented; (iv) Clinical Groups and SEER groups; (v) Sample Source; (vi) Technique. Additional options to subset the data are again (vii) Find CNAs by Gene or Region and the possible restriction to samples with (viii) Clinical Data.

After the "sample selection" step, the resource advances to the "Data Selection and Visualization Options" page (Figure 1B). The purpose of this interface is to selectively customize plot options and parameters. Of the parameters, we here only want to mention the possibility to restrict CNAs only to such of a given size range (e.g. excluding all whole-chromosome changes) with a "Segment Size Filter"; the labeling of regions of interest using a gene selector or free Golden Path coordinate entry in "Mark Region or Gene Locus"; the adjustment of histogram plot parameters such as plot size, range and labeling in "Histogram Plot Options" as well as the type of data clustering method and sample display. For complex data sets, one helpful option is the "Group Analysis" feature. As example, when used in the esophagus ca. data set, setting the value to "ICDMORPHOLOGY" and requiring a minimal group number of 50 will produce additional CNA histograms for both adenocarcinomas and squamous cell carcinomas of the esophagus, as well as a small heatmap presenting the CNA frequencies of those groups side-by-side (Supplementary Figure S2).

Another option in this section is to generate both group specific as also locus related Kaplan-Meier plots for samples with follow-up data. This option may be used to explore the possible association of regional CNAs and clinical risk, as a preliminary step for proper gene specific risk attribution.

Although at this time we do not attempt to provide the infrastructure for hosting of user generated analysis results, a basic framework is given to generate named, temporary directories. This provides users with the option to define a protected directory in which to save intermediate analysis results. Detailed instructions on how to navigate the website are available in the "FAQ & Guide" page: http://www.progenetix.org/cgi-bin/reader.cgi?project=progenetix&tags_m=guide.

**Pre-defined data organization**

Additionally to the *ab novo* sample selection, the contents of Progenetix can be browsed through pre-defined cancer groups as classified by ICD-O-3 coding system, tumor site, clinico-pathological entities and SEER (44), respectively. These groups allow users to quickly access data for a specific cancer type. At the moment, the most comprehensive and detailed standard for cancer classification is ICD-O 3 (International Classification of Diseases for Oncology, 3rd Edition) (42). It is a coding system developed by the World

Health Organization (WHO) and describes entities based on tumor site (topography) and histology (morphology). In total, 363 ICD-O 3 entities are recorded in Progenetix, and serve as the primary classifier for most analyses. The second standard classifier is based on the ICD-O topography code, i.e. the tumor's site (Table 1). According to this system, all database records are categorized to 80 loci; however, this also mirrors the fact that for many samples the assignment granularity is limited (e.g. C069 "mouth, NOS" instead of e.g. C062 "retromolar area"), and/or that sample sizes for some specific loci are limited, leading to assignment to the more general category. The third classification system represents clinico-pathological entity groups; essentially, this system captures the approximation of standard diagnostic assignments (e.g. "Carcinomas: Colorectal Adenoca." for all types of adeno-carcinomas with location in large intestine or rectum). So far, 83 diagnostic groups have been defined in Progenetix. The last system is established by the Surveillance, Epidemiology and End Results, a public resource for cancer statistics or cancer surveillance methods. According to Progenetix data, 70 SEER cancer groups are represented in the database.

## DATA ANALYSIS TOOLS

**Data visualization and exploration**

To exemplify the new data visualization interface and the online analysis tools of Progenetix, Figure 2 illustrates the results of processing the pinealis region tumor data, represented by 27 samples. The first panel of Figure 2 is a circos-style (45) plot that shows the frequency and concurrence of all copy number alterations found in pinealis region tumors (Figure 2A). The chromosome ideograms are displayed with cytobands, oriented from the p-arm of chromosome 1 to the q-arm of chromosome 22 in a clockwise direction with centromeres indicated as purple bands. The frequency of genomic gains and losses among the 27 samples is presented in the inner circle by yellow and blue areas, respectively. If the dataset is of low-complexity, there will be ribbons representing the connections between all concurrent in-case imbalances. In the chromosome ideogram (Figure 2B) yellow and blue bars with percentage label on the right and left side of the chromosome represent the frequency of gains and losses, respectively. The histogram shows the CNA frequencies throughout the genome for building a profile of chromosomal rearrangement hotspots (Figure 2C). This figure may be particularly helpful in the genome-wide identification of copy number imbalance peaks, which may point to genomic loci harboring cancer related genes. Sample specific CNAs are displayed in the "matrix plot" panel, with hierarchical clustering applied as selected (Figure 2D). In this case, color labels point to different values for PubMed ID, ICD morphology and topography and will be detailed as "mouseover" event when opening an SVG version of the image.

If several cancer types are selected for this analysis, differences in CNA patterns among different cancer types can usually be observed. In the frequency matrix, a black-to-yellow gradient is used to indicate the frequency of gains from lowest to highest, while the frequency of losses is given by the gradient from black-to-blue (Figure 2E). This matrix is particularly helpful when comparing CNA profiles among several cancer groups with the intuitive and global view of regional hotspots. The last section is the "Sample Data", listing the summary of each sample that included in this analysis (Figure 2F). Clicking on an individual record will lead to a page that provides detailed information of the single sample, as well as the graphical representation of the samples CNAs.

**Gene CNA frequencies**

Cancer related genes may play crucial roles in cancer development, and can be classified into the two basic types of oncogenes and tumor suppressor genes (8,9). For a number of oncogenes (e.g. *ERBB2*, *MYCN*, *REL*, *CDK1*), functional activation based on a "dose-effect" due to genomic copy number gains has been shown. Conversely, tumor suppressors are characterized by reduced activity and frequently targets of genomic deletions (e.g. *TP53*, *CDKN2A/B*, *RB*, *APC*). When exploring a candidate cancer related gene, one of the interesting questions is the frequency of copy number abnormalities involving the gene's locus in different cancer types. In this new release of Progenetix, we provide an online tool in the page "Gene CNA Frequencies", to help investigate cancer gene status based on the large amount of tumor samples. Users can search for single or combined imbalances by selecting gene names from the auto-complete list, or manually specify loci (cytogenetic bands or "Golden Path" coordinates) and types of the changes of interest. Furthermore, CNAs can be limited to focal events (e.g. smaller than 5Mb), to increase the specificity of the required change through the exclusion of large CNAs affecting multiple possible targets. Please note, that this option is somewhat limited due to the limited spatial resolution of the cCGH data sets (13) included in Progenetix. The "Minimal Case Number" field provides a threshold to improve the reliability of the query results through the optional removal of cancer entities with limited sample number. The result page indicates a list with subset specific data: (i) the relative number and percentage of samples with the hit; (ii) a score value that weighs the hit frequency by the subset's overall genome complexity (hit frequency divided by the average genome CNA coverage of the subset's samples). Here, higher complexity samples will have a lower score, due to the overall high probability to display a hit in any given region. The returned samples can be used for further processing and visualization.

## OTHER IMPROVEMENTS

**User file processing**

We provide a "User File Processing" interface for users to take advantage of the online tools by uploading their private data. These data can be tab-delimited text file, a pre-processed JSON data file (e.g. from a previous Progenetix analysis run) or one of a number of segmentation file types as generated by genomic array analysis procedures. Depending on the file type selected, CNAs may be annotated either using Golden Path coordinates and CNA type or value/threshold combinations, or be provided in a cytogenetic annotation format (ISCN "ish cgh" style). Uploaded data is processed into the standard internal BSON format, and can be retrieved as JSON file for storage or directly be processed in the standard visualization pipeline described before.

Although we focus on human cancer genome data, the online visualization and exploration tools can be applied to other species. Recently, the *Danio rerio* genome coordinates have been added to the tool to allow for zebrafish genome data processing. This interface is easily extendable upon user request.

Additionally to the general curation and representation of cancer CNA data, Progenetix is now being used as hosting framework for disease specific project data. We have recently started to provide the backbone and data interface for two collaborative projects. The DIPG project focuses on diffuse intrinsic pontine gliomas (46,47). It aims to provide a central resource for researchers to investigate genome-wide profiling data from these devastating childhood brain tumors as well as from other rare, aggressive pediatric gliomas (48,49). A significant amount of data have been submitted into the database by collaborators and supporters and was integrated with publicly-available data sets to provide a systematic review and meta-analysis of these diseases. The other current project is aimed at cutaneous T-cell lymphomas and related, inflammatory skin diseases.

**Database implementation, formats and API**

The Progenetix site runs on a MongoDB backend in a Unix environment (Apple Mac OS X). Data is stored in a sample specific manner, with pre-computed CNA status interval data (1Mb resolution) and sample-specific segments (resolution only limited by original analysis technique). For data downloads, JSON data files are provided, as well as tab-delimited data formats for CNA status matrices and sample annotation files. With a general availability since February 2013, Progenetix now provides a query based API for programmatic access and image generation. Documentation including query parameters and examples as well as relevant updates regarding query constructs and output formats can be accessed through the documentation at http://www.progenetix.org/cgi-bin/reader.cgi?tagsearch_m=api.

## CONCLUSION

Progenetix is a comprehensive, curated oncogenomic database that provides copy number aberration data to the human cancer research community. Over the past 12 years, the database has undergone an extensive expansion and significant qualitative enhancements. Particularly, the database has made the transition from a "cytogenetic" resource based on cancer cytogenetic data to an integrated resource incorporating cancer genome data from increasing variety of genome analysis techniques. Likewise, many ideas of the user interface improvements and data analysis tools have been implemented based on suggestions from users. While providing genomic aberration data from the largest range of cancer entities available, in the future we will especially focus on an extension of the data model and improved inclusion of associated clinical information, as well as a tighter integration with online repositories and array repositories (e.g. http://www.arraymap.org).

## ACKNOWLEDGEMENTS

Particularly thanks are due to Chris Jones and his group at Cancer Research UK for ongoing collaborative development of the diffuse intrinsic pontine glioma project. We would also like to acknowledge the Progenetix users who have provided valuable feedback.

## FUNDING

This work was supported in part by the University of Zurich's Research Priority Program Systems Biology. HC is supported through a personal grant from the China Scholarship Council.

*Conflict of interest statement.* None declared.

Table 1. The full complement of Progenetix data summarized by cancer loc

| Cancer loci | cCGH | aCGH | Publications |
|---|---|---|---|
| hematopoietic and reticuloendothelial systems | 2580 | 2689 | 170 |
| lymph nodes | 1181 | 1164 | 61 |
| breast | 1259 | 1012 | 65 |
| cerebellum | 674 | 765 | 59 |
| brain, NOS* | 845 | 497 | 78 |
| cerebrum | 452 | 749 | 49 |
| liver | 1054 | 126 | 56 |
| stomach | 977 | 178 | 46 |
| skin | 889 | 184 | 46 |
| connective and soft tissue, NOS | 1001 | 57 | 63 |
| kidney | 723 | 295 | 40 |
| large intestine, excl. rectum and rectosigmoid junction | 572 | 429 | 51 |
| ovary | 587 | 146 | 27 |
| prostate gland | 640 | 95 | 20 |
| lung and bronchus | 441 | 258 | 28 |
| nervous system, NOS | 421 | 246 | 18 |
| urinary bladder | 364 | 223 | 14 |
| cervix uteri | 411 | 118 | 17 |
| peripheral nervs incl. Autonomous | 290 | 233 | 24 |
| esophagus | 426 | 28 | 22 |
| pancreas | 376 | 50 | 17 |
| thyroid gland | 385 | 19 | 17 |
| pleura | 311 | 72 | 24 |
| bones and joints | 325 | 25 | 21 |
| Spleen | 56 | 222 | 11 |
| other | 3677 | 845 | 237 |
| Total | 20917 | 10725 | 1006 |

* NOS, not otherwise specified

# FIGURES

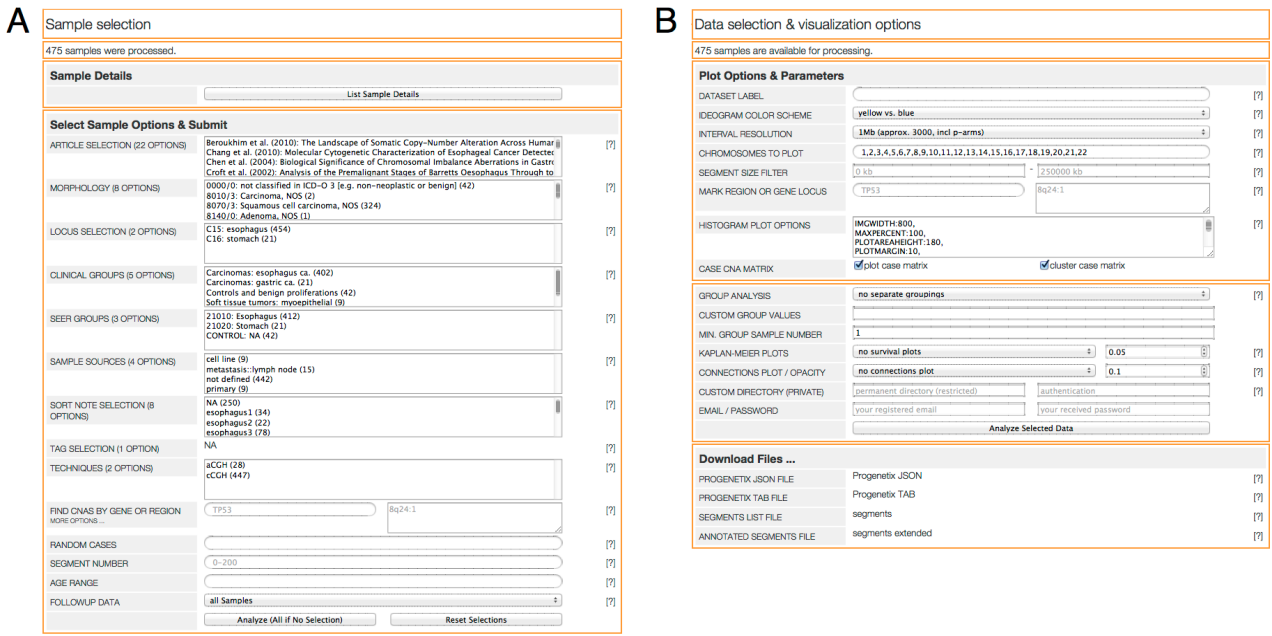

Figure 1. A screenshot of the data selection page showing the new layout of the search fields. (A) Sample selection. (B) Data selection and visualization options. In this example, 475 records were returned when searching for the keyword "esophagus".

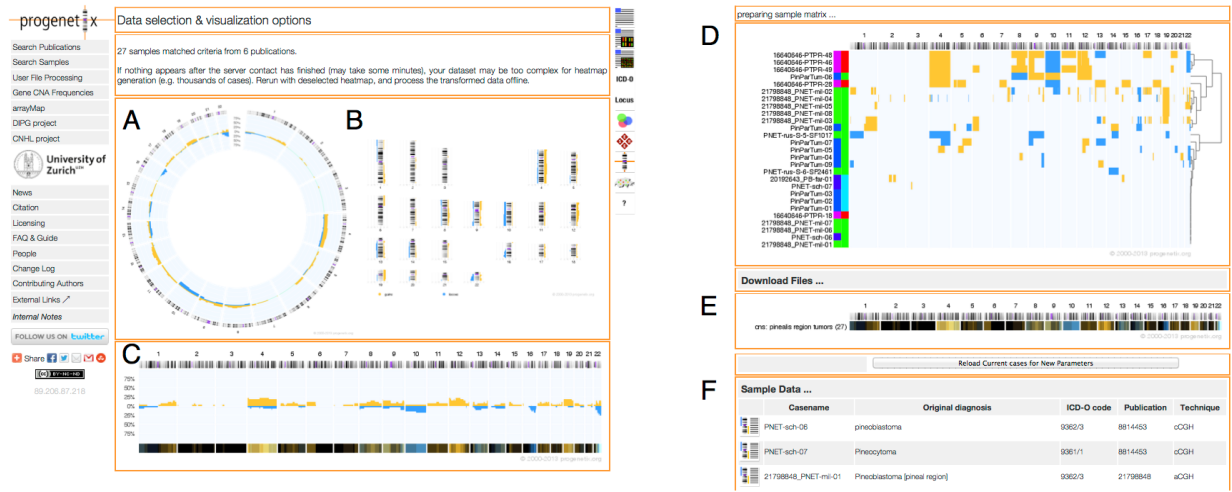

Figure 2. Results page of online statistical analysis tools. In this example, 27 pinealis region tumor samples have been selected. The functional blocks illustrate the website's new functionality. (A) Circular CNA plot. (B) Histogram of genomic imbalances. (C) Frequency plot across the genome. (D) Matrix plot showing case-level copy number aberrations. (E) Frequency matrix. (F) Sample details and links to single case visualization. See text for further details.

## Supplementary Figures

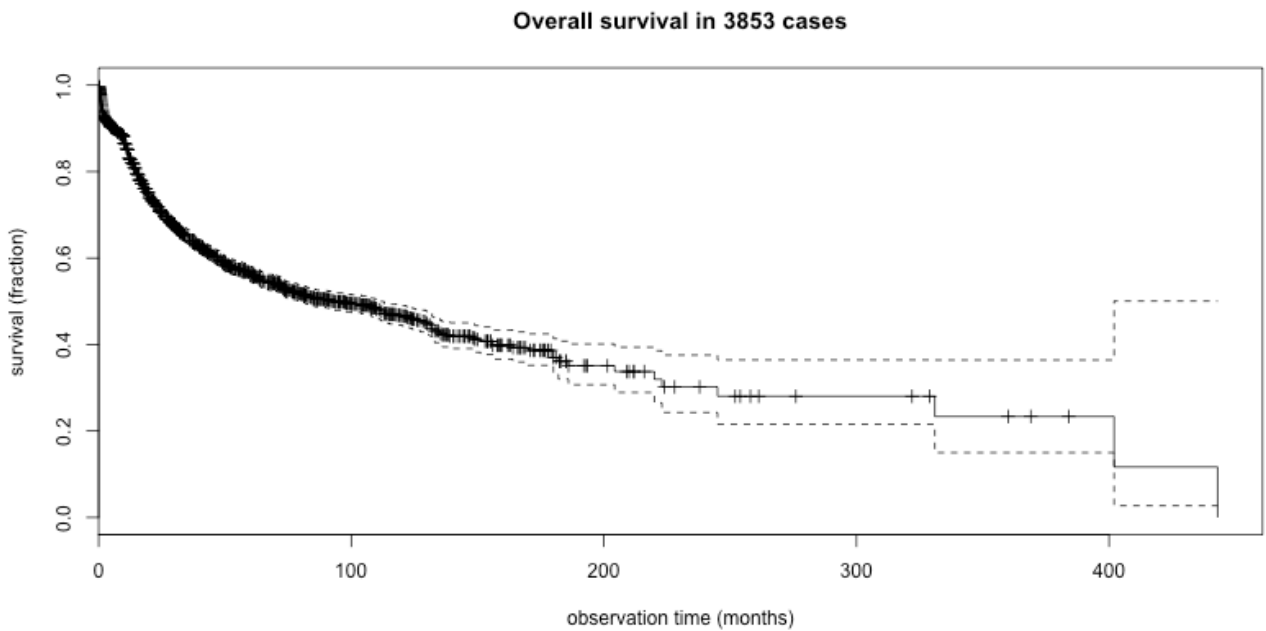

Figure S1. Overall survival curve in 3853 Progenetix cases.

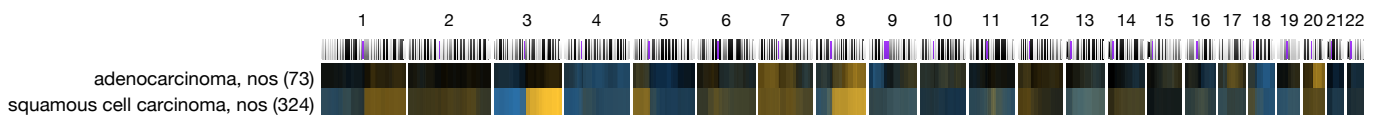

Figure S2. A heatmap of two subtypes of esophagus carcinoma. The numbers in brackets are number of cases. Chromosome 1 to 22 and bands are presented on the top of the figure.